\documentstyle[12pt]{article}

\begin{document}
% Define sanserif font
%\font\sansf=cmss12
\font\ticp=cmcsc10

\begin{titlepage}

\rightline{DAMTP-R94/61}
\rightline{Phys. Rev {\bf D52.4} (1995)}
\vspace{0.1in}
\LARGE
\center{Entropy in the RST Model }
\Large
\vspace{0.2in}
\center{}
\center{Justin D. Hayward\footnote{E-mail address:
J.D.Hayward@damtp.cam.ac.uk}}
\vspace{0.2in}
\large
\center{\em Department of Applied Mathematics and Theoretical
 Physics,
\\ University of Cambridge,
\\  Silver Street, Cambridge CB3 9EW, U.K.}
\vspace{0.2in}
\center{December 1994}

\small\center{ }
\vspace{0.3in}
%\newpage

\begin{abstract}
The RST Model is given boundary term and Z-field
so that it is well-posed and local. The Euclidean method is
described for general theory and used to calculate the RST
intrinsic entropy.
 The evolution of this entropy
for the shockwave solutions is found and obeys a second law.

\vspace{0.3truecm}
\end{abstract}

\setcounter{footnote}{0}
\end{titlepage}

\section{Introduction}
\setcounter{equation}{0}
A notable development in the ongoing study of the black hole
evaporation
problem\cite{REVS} is that there is now a two dimensional
model which
admits physically sensible evaporating black hole solutions
\cite{RSTB}, at least
for much of the process.

In\cite{SHM}, it was shown that if the radiation in the scalar
degrees of freedom
which
produce a black-hole like object, has
positive energy, there must be a global event horizon or a naked
singularity.
 This was a general argument which did
not depend upon the precise fine-tuning of the model: only the
general form of the global structure was determined.
If a spacetime has a global event horizon or a naked singularity,
one can say through
reasonable qualitative statements that the former case implies
loss of quantum
coherence, while the latter would lead to
an even worse breakdown of predictability. Whilst such qualitative
arguments are
convincing, at least at the semi-classical level, (a consideration
of
superselection sectors in a third quantised framework\cite{SP} may
 lead to other
possibilities), one would
like to prepare explicit calculations which show that some physical
quantity
changes during the
evolution of a spacetime in such a way that it is clearly seen
whether there is loss of
information. The quantity usually considered is the entropy.
A good question then is how to relate the entropy changes to
information loss\cite{BekA,BekB}.
 This will be considered elsewhere.

 Here a particular model is considered,
and the behaviour of its quantitative entropy.
The Euclidean method for calculating the entropy is described,
and compared with that of another method. The
change in entropy during
the lifetime of the class of black holes formed by the collapse
 of a shock wave is then calculated.
This expression shows that the formation and evaporation of
a black hole causes the entropy to increase for all positive energy
incoming
pulses. One can also show that the entropy is always increasing,
 that is,
 these
solutions obey a second law.

\section{Local Well-Posed RST Model}
\setcounter{equation}{0}

In this section, a model which has been discussed many times in
the
literature\cite{CGHS} is discussed, but here the necessary boundary
term for the
variational
problem to be well-posed and for the thermodynamics to be derived
using the
Euclidean formalism is included . The quantum effects of the minimal
 scalars are
represented by introducing an independent scalar field which mimics
 the
trace anomaly of
the
matter fields, and also makes the effective action local, thus
simplifying the
 calculation
of the field equations.
The effective action needed is the following:-

\begin{equation}
I = \frac 1 {2\pi}\int d^2 x \sqrt{\pm g}
\Big[ R\tilde\chi+4({\nabla \phi}^2+\lambda^2)e^{-2\phi}
-\frac {\kappa} 4 {\nabla Z}^2-\frac 1 2 \sum_{i=1}^N(\nabla f_i)^2
\Big]
-\frac 1 {\pi} \int d \Sigma \sqrt{\pm h} K\tilde\chi   \label{eq:SM}
\end{equation}
where $\tilde\chi=e^{-2\phi}-\frac {\kappa} 2 (\phi-Z).$

The fields present are the metric, $g_{\mu\nu}$, the dilaton,
$\phi$, the scalar field $Z$, and the classical minimal scalars.
The pair of terms involving Z in the volume term of this action have
replaced
the
usual Polyakov\cite{POL} term
\begin{equation}
\frac {\kappa} 4 \int d^2x \sqrt{-g(x')} G(x,x')R(x') \label{eq:POLT}
\end{equation}
This is multiplied by R and integrated in the action(\ref{eq:SM})
 and arises from the matter fields' contribution to the
associated path integral. The trace anomaly of the Z-scalar field is
that of the N minimal scalars.
 There is a boundary term
\footnote
{Note that the Noether charge method of calculating the entropy due
to
Wald\cite{WALD}
 does not mention the boundary term, which is essential in the
Euclidean derivation. The equivalence of the methods is described
later in
the same paper. }
which defines the variational problem properly.

The equations of motion are
\begin{eqnarray}
& & R_{\mu\nu}(1-\frac {\kappa} 4 e^{2\phi})
+2 \nabla_{\mu} \nabla_{\nu} \phi(1+\frac 1 4 \kappa e^{2\phi})
\nonumber \\
&=&\frac {\kappa} 4 e^{2\phi}(2\nabla_{\mu}\nabla_{\nu} Z
+\nabla_{\mu} Z \nabla_{\nu} Z-
\frac 1 2  g_{\mu\nu}(\nabla Z)^2 )  \nonumber \\
& & +\frac 1 2  e^{2\phi}(\nabla_{\mu} f_{i} \nabla_{\nu} f_{i}-
\frac 1 2  g_{\mu\nu}(\nabla f_{i})^2)  \label{eq:eBG}
\end{eqnarray}

\begin{equation}
R(1+\frac {\kappa} 4 e^{2\phi})=4 (\nabla \phi)^2 -4 {\nabla}^2 \phi
- 4\lambda^2        \label{eq:eBDIL}
\end{equation}

\begin{equation}
{\nabla}^2 Z + R = 0       \label{eq:BZ}
\end{equation}
and the minimal scalars obey the wave equation.
More information about the form of the Z field can be deduced from
these equations.
If one takes the trace of the gravitational field equation, and uses
(\ref{eq:BZ}),
 one finds that $R=-2{\nabla}^2\phi$.
But eqn.(\ref{eq:BZ}) then implies that
\begin{equation}
Z=2\phi + \xi      \label{eq:ZP}
\end{equation}
where $\xi$ is a solution of the wave equation.
If one were to introduce the conformal factor $\rho$, which appears
in the
conformal metric in the form
\begin{equation}
ds^2=-e^{2\rho}\eta_{\mu\nu}dx^{\mu}dx^{\nu}  \label{eq:CM}
\end{equation}
then one has $R=-2{\nabla}^2 \rho$
so that the Z field is clearly related to the conformal factor by
\begin{equation}
Z=2\rho +\eta   \label{eq:ZQ}
\end{equation}
where $\eta$ is another solution of the wave equation. Also,
evidently,
the conformal factor and the dilaton have to differ only by such a
solution,
 which depends on gauge.
\section{Semi-Classical Euclidean Entropy }
About twenty years ago, a close mathematical analogy between
thermodynamics
and the mechanics of
black holes was noticed by Bekenstein\cite{Bek}. The calculations of
 Hawking
\cite{SHA} then put
this analogy on a safer footing by supplying constants of
 proportionality.
Hawking and Gibbons\cite{GH} showed how to use the Euclidean
continuation of a
 spacetime to derive its entropy.

One first writes down the amplitude for  a field configuration
$\phi_1$ at
time $t_1$ to another field configuration $\phi_2$ at time $t_2$.
This is
\begin{equation}
\langle {\phi_2,t_2|\phi_{1},t_1} \rangle=\langle
{\phi_2|e^{-iH(t_2-t_1)}|\phi_1} \rangle
= \int D[\phi] e^{iI(\phi)}    \label{eq:Pa}
\end{equation}
If one now lets $-i\beta=t_2-t_1$ be purely imaginary, identifies
the
initial and final field configurations, and sums over all possible
states,
the result is
 the partition function for the field on the Euclidean spacetime,
at temperature $T=\beta^{-1},$
\begin{equation}
\sum_{n}\langle\phi_{n}|e^{-\beta H}|\phi_{n}\rangle=
\int D[\phi] \exp[-i\hat I (\phi)]=Z  \label{eq:Paa}
\end{equation}

Once the identification of the thermodynamic partition function has
been made,
 one
can proceed as usual to derive other thermodynamical quantities, such
 as the
entropy.
This is done by finding the
dominant contribution to the partition function and then substituting
 into the
usual formula for
 the entropy:
\begin{equation}
S=(\beta\frac {\partial} {\partial\beta}-1 )\log Z = \beta
\frac {\partial I} {\partial\beta}-I   \label{eq:SB}
\end{equation}

Consider a Euclidean solution with an inner boundary just outside the
 horizon.
 One could
identify this spacetime with any period. Evaluating the action shows
that
it is proportional
to the period. Thus from (\ref{eq:SB})the entropy will be zero. If
however the
period,$\beta$,is
taken to be that value which removes the conical singularity at the
origin,
 one
can remove the inner boundary. This changes the topology and reduces
the
 action
by the surface term at the origin evaluated such that the conical
 singularity
is
removed. This quantity is independent of $\beta$. One can
 interpret
this reduction in the action as corresponding to an increase in the
 partition
function which in turn corresponds to an increase in the entropy.
 This will be
equal to the term described above, as can be seen from (\ref{eq:SB}).
 This point will be
discussed further in \cite{HH}.

In this derivation, one effectively considers the entropy
defined at a given instant, making the approximation that the
evaporation is quasi-static. The Z-field accounts for the
dynamics of the solution and appears in the formula for the entropy.
One is in effect comparing the real  solution with a sequence static
black holes with the same values of $\phi$ and
 Z on
the horizon.

In the case of the RST model  one finds for the on-shell action,
\begin{equation}
I=-\frac 1 {\pi} \int d \Sigma \sqrt{h}  \Big(K\tilde\chi
-2 e^{-2\phi}\nabla_{n}\phi+\frac {\kappa\phi\nabla_{n}\phi} 4
-\frac {\kappa} 2
((\xi+\phi)\nabla_{n}\phi+
\frac 1 {2\kappa} f_{i}\nabla_{n} f_{i})\Big) \label{eq:ION}
\end{equation}
where $\tilde\chi=e^{-2\phi}-\frac {\kappa} 2 (\phi-Z).$

There are inner and outer boundaries, but only the inner one
contributes.
The terms involving K are those which have come from the surface term,
 and will
be all those which contribute to the entropy. This is because the
1-volume of the inner boundary tends to zero, and the other terms
 are
regular there.   As explained above,
 one takes the period such that the horizon
point is regular, which is necessary for a sensible Euclidean
spacetime
\footnote{for staticity and regularity $\nabla_{n}\rho |_{h}=0,$}.

To evaluate (\ref{eq:ION}), suitable coordinates in Euclidean
signature are:

\begin{equation}
ds^2=e^{2\rho (r)}(dr^2+r^2 d\theta ^2)  \label{eq:EUCMET}
\end{equation}

The coordinate transformations needed to transform from Kruskal
conformal
coordinates
  to (\ref{eq:EUCMET}) are
$x^{+}=re^{i\theta}, -x^{-}=re^{-i\theta}$, so that
 $r^2=-x^{+}x^{-}$ and $\theta=\frac 1 {2i}\log -\frac {x^{+}}
{x_{-}}.$ In these coordinates, $\sqrt{h} = r e^{\rho}$ and the trace
of the second fundamental form K for a circular boundary is given by
\begin{equation}
K=-\frac {1} {re^{\rho (r)}} (1+r\frac {d\rho} {dr}).  \label{eq:K}
\end{equation}

  One finds that the action is
\begin{equation}
-\frac 1 {\pi} \int d \Sigma \sqrt{h} K\tilde\chi  \label{eq:LOOP}
\end{equation}

Since the period is identified with $2\pi$,
$\sqrt{h} K= -1,$ and the action becomes
\begin{equation}
I=2(e^{-2\phi_h}-\frac 1 2 \kappa\phi_h+\frac 1 2 \kappa Z_h)
\label{eq:LOOPP}
\end{equation}

This, as explained above, is the entropy. This is the result one
obtains
using the Noether charge
method. This was expected as the two methods, though apparently very
dissimilar, are equivalent\cite{WALD} as was mentioned before.

As the black hole loses mass, the value of the dilaton on the horizon
decreases\cite{W}. The dilaton terms in the entropy correspond to
the  classical and
quantum terms due to the black hole itself. The Z-field term includes
the
 effects of the radiation. As Z increases on the horizon, it
compensates
for the reduction in internal entropy. How the radiation term
increases
is a measure of the
inefficiency of information retrieval from the black hole. For
thermal
radiation, its rate of increase should be a maximum, {\it i.e. } any
 other kind
of radiation
should give a slower rate of increase. If the difference between
these
rates is large enough, then it might be possible that all the
information
which went into producing the black hole would be retrieved, albeit
in a
form very difficult to collect. In terms the entropy, this would
be
described by the case where the radiation terms exactly cancelled
the
changes due to the internal contributions.

It is clear that in the exact RST shockwave solutions, the radiation
 is no
longer exactly
thermal\cite{MATHUR}.  Thus, they must radiate {\it some}
information about the initial state, if only a bit.
 Since our formula (\ref{eq:LOOPP}) is derived
using the quasi-static approximation, and the evolving RST solutions
 are not
strictly quasi-static, the statements one can make about the entropy
 changes
found in the following chapter are really only guidelines. One should
 rather
 consider a series of static
black holes with the same values of dilaton and Z-field on the
horizon, but
with decreasing mass.

\section{Entropy Evaluation for RST Shockwaves}
\setcounter{equation}{0}

We now calculate the entropy for solutions describing the evolution of
 spacetime
 after a  shockwave of positive energy, m, is sent in from past right
infinity. These form black holes which evaporate away to nothing in a
finite lifetime.
This is done in the conformal gauge, and in double null coordinates,
 {\it i.e.} the metric is
\begin{equation}
ds^2=-e^{2\rho}dx^{+}dx^{-}   \label{eq:CMI}
\end{equation}
Since the Z field obeys (\ref{eq:ZQ}), where $\eta$ is a solution of
the
 wave equation, the field
equations in terms of the new variables $\Omega$ and $\chi$ used by
RST,
 are
unchanged. It seems not possible to write the local RST model in
variables which reduce it to a Liouville-like model.
 Therefore, the usual transformation(see \cite{RSTB}) are used.
 The definitions of the new variables are
\begin{equation}
\Omega=\frac 1 {2\sqrt{\kappa}}(2e^{-2\phi}+\kappa\phi)
 \label{eq:Om}
\end{equation}
\begin{equation}
\chi=\frac 1 {2\sqrt{\kappa}}(2e^{-2\phi}-\kappa\phi+2\kappa\rho)
\label{eq:CHI}
\end{equation}
and the volume part of the action becomes
\begin{equation}
S={1\over\pi}\int
d^2x\Big(-\partial_+ \chi\partial_-\chi
+\partial_+\Omega\partial_-\Omega
+\lambda^2\exp\left({2\over\sqrt\kappa}(\chi -\Omega
)\right)
+{1\over2}\sum_{i=1}^N\partial_+ f_i\partial_- f_i\Big)
\label{eq:SMOM}
\end{equation}
The field equations for the action
(\ref{eq:SM}) reduce neatly to the
 following form
\begin{equation}
\partial_+\partial_- (\chi - \Omega)=0   \label{eq:} \label{eq:FEI}
\end{equation}
\begin{equation}
\partial_+\partial_-( \chi + \Omega)=-{2\lambda ^2\over \sqrt
\kappa}\exp \left({2\over \sqrt \kappa}(\chi - \Omega)\right)
\label{eq:FEII}
\end{equation}
with constraint equations
\begin{equation}
\partial_{\pm}\Omega\partial_{\pm}\Omega-
\partial_{\pm}\chi\partial_{\pm}\chi+
\sqrt{\kappa}{\partial_{\pm}}^2\chi
+\frac 1 2 \partial_{\pm} f_{i}\partial_{\pm} f_{i}=\kappa t_{\pm}
\label{eq:CONI} \end{equation}
 where the functions $t_{\pm}$ are
chosen so that for a given solution to the field equations the
constraint
equations
are satisfied.  The last term on the left hand side is
summed over N matter fields.

One can fix the remaining coordinate freedom left in the conformal
 gauge by
setting
$\Omega=\chi$, which is the Kruskal Gauge, $\rho=\phi$ in terms of
the old
variables.
The general solutions for infalling matter are
\begin{equation}
\Omega ={-\frac {\lambda ^2} {\sqrt\kappa}} x^{+}(x^{-}+ P_{+}(x^+))
+\frac 1 {\lambda\sqrt{\kappa}} M(x^+) -\frac {\sqrt{\kappa}} 4
\log(-\lambda ^2x^+x^-)   \label{eq:GEN}
\end{equation}
where M is the $x^+$-dependent Kruskal energy-momentum tensor
\begin{equation}
\lambda\int_{0}^{x^+} dy y T_{++}(y)   \label{eq:MK}
\end{equation}
and $P_{+}$ is the total incoming momentum up to the advanced time
$x^+$
given by
\begin{equation}
\int_{0}^{x^+} dy  T_{++}(y)   \label{eq:PK}
\end{equation}
Solutions for which
\begin{equation}
T_{++}=\frac 1 2 \partial_+ f_{i}\partial_+ f_{i}=\frac m {\lambda
x_{0}^{+}} \delta(x^{+}-x_{0}^{+})     \label{eq:SH}
\end{equation}
are considered.
These are incoming shock waves at $x^{+}=x_{0}^{+}$, with total
energy m.

In order to find S, one must fix Z by imposing initial conditions
along a
Cauchy surface. This is done in such a way that corresponds to no
incoming
particles from right infinity except for the shockwave, and none in
spacetime before the shock. These are the same conditions that were
applied to
the Z-field introduced in \cite{SHM}, which are shown in FIG.1; in
this case, however, although  the
energy-momentum tensor still mimics the trace anomaly of the minimal
scalars,
 one is
now of course treating a particular model rather than the general
form in two
dimensions.

\setlength{\unitlength}{0.0100in}%
\hspace{1.6in}
\begin{picture}(200,300)(300,440)
\put(380, 595){\line( 1, 1){72}}
\put(380, 595){\line( -1, -1){60}}
\thicklines
\put(320,600){\line( 0,-1){160}}
\put(320,440){\line( 1, 1){180}}
\put(500,620){\line(-1, 1){120}}
\put(380,740){\line( 0,-1){140}}
\put(320,600){\line( 1, 0){ 60}}
\put(320,600){\line( 1, 0){ 60}}
\put(320,595){\line( 1, 0){ 60}}
\put(380,595){\line( 0, 1){145}}
\put(395,520){\vector(-1, 1){ 75}}
\multiput(380,595)(19.92875,19.92875){4}{\line( 1, 1){ 10.214}}
\multiput(380,595)(-24.89312,-24.89312){3}{\line(-1,-1){ 10.214}}
\put(425,525){\vector(-1, 0){ 35}}
\put(440,530){\vector( 0, 1){ 30}}
\put(340,515){\makebox(0,0)[lb]
{\raisebox{0pt}[0pt][0pt]{ I}}}
\put(420,590){\makebox(0,0)[lb]
{\raisebox{0pt}[0pt][0pt]{ II}}}
\put(395,655){\makebox(0,0)[lb]
{\raisebox{0pt}[0pt][0pt]{ III}}}
\put(370,550){\makebox(0,0)[lb]
{\raisebox{0pt}[0pt][0pt]{ matter}}}
\put(375,535){\makebox(0,0)[lb]
{\raisebox{0pt}[0pt][0pt]{      wave}}}
\put(430,520){\makebox(0,0)[lb]
{\raisebox{0pt}[0pt][0pt]{Z=0}}}
\put(290,515){\makebox(0,0)[lb]
{\raisebox{0pt}[0pt][0pt]{r=0}}}
\put(350,650){\makebox(0,0)[lb]
{\raisebox{0pt}[0pt][0pt]{r=0}}}
\end{picture}
%\twlrm

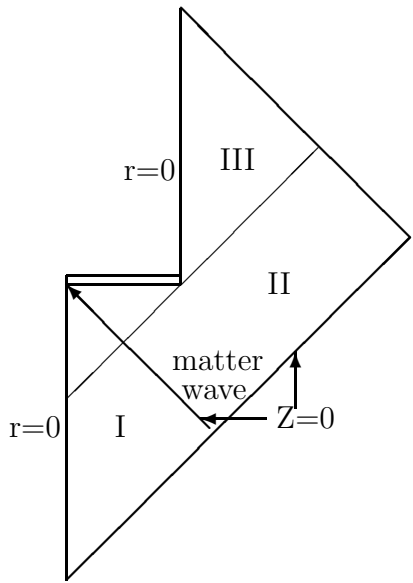
\centerline{Figure 1: Penrose diagram of shockwave solutions}
\vspace{0.2in}

Since
\begin{equation}
Z=2\phi + \xi,
\end{equation}
Comparing the constraint equations derived from (\ref{eq:eBG}),
 in the Kruskal gauge,
with equation (\ref{eq:CONI}) shows that one must identify
\begin{equation}
t_{\pm}=-(\partial_{\pm}^{2}\xi+\partial_{\pm}\xi\partial_{\pm}\xi).
\label{eq:tt} \end{equation}
The work can now be connected to that of \cite{MY} by identifying the
 function
$\xi=\xi_{+}+\xi{-}=w_{+}+w_{-}.$ In that paper, $w_{\pm}$ were the
 functions
which are added to the Kruskal conformal factor to obtain the
conformal
factor in asymptotically flat coordinates relative to which one
defines the vacuum.

The function $\xi$ can be deduced if Z is specified on two null
lines.

The solution before the shock is the linear dilaton
\begin{equation}
\phi=-\frac 1 2 \log(-\lambda ^2 x^{+}x^{-})  \label{eq:LDV}
\end{equation}
and the dilaton is continuous everywhere, so one knows that on the
shockwave
itself
at $x^{+}=x_{0}^{+}$ the dilaton is still the above.
 And on right infinity, the solution also tends to this form.
Setting $Z=0$ on the shockwave line  $x^{+}=x_{0}^{+}$ and on right
past
infinity,
\begin{equation}
\xi=\log(-\lambda ^2x^{+}x^{-})   \label{eq:XI}
\end{equation}
The entropy for this scenario can now be evaluated.
Using the relation for Z in terms of $\phi$ (\ref{eq:ZP}) and the
definition of  $\Omega$,
the entropy formula can be rewritten,
\begin{equation}
S=2\sqrt{\kappa}\Omega_{h}+k\xi_{h}   \label{eq:EUCL}
\end{equation}
Note that the function $\xi$ can be related to the functions of
integration
$t_{\pm}$
 First take `h' to be the apparent horizon. One has the solution for
$\Omega$
using (\ref{eq:GEN}) and (\ref{eq:SH}). $\xi$ is known from
(\ref{eq:XI}). The equation of
the apparent horizon is thus needed, which is also straight forward.
The apparent horizon is
defined by $\partial_+\Omega=0$. This gives, in terms of the chosen
Kruskal
 coordinates,
\begin{equation}
4x_{h}^+(x_{h}^- +\frac m {\lambda ^3 x_{0}^{+}})=-\kappa\lambda ^2
 \label{eq:AH}
\end{equation}
To express the total entropy change from start to finish, the
coordinates of the endpoint
are also required. This is found simply by solving simultaneously
 with (\ref{eq:AH})  the
equation for the singularity (which is on the line where
$\frac {d\Omega} {d\phi}=0$
which gives $\phi_{cr}=-\frac 1 2 \log\frac \kappa 4$ and so
$\Omega=\Omega_{cr}=\frac {\sqrt\kappa} 4 (1-\log\frac {\kappa} 4)$).

This yields for the endpoint
\begin{equation}
x_{e}^{+}=\frac {\kappa \lambda x_{0}^{+}} {4m} (e^{\frac {4m}
{\kappa\lambda}}-1)  \label{eq:ENDP}
\end{equation}
\begin{equation}
x_{e}^{-}=-\frac m {\lambda ^3 x_{0}^{+}} \frac {e^{\frac {4m}
{\kappa\lambda}}}
{e^{\frac {4m} {\kappa\lambda}}-1}    \label{eq:ENDM}
\end{equation}
The above agrees with \cite{RSTB}.
Knowing the beginning and endpoints of the apparent horizon one can
substitute into the entropy formula to find the overall change.
This turns out
to be
\begin{equation}
\Delta S=S(x_{e}^{+},x_{e}^{-})-S(x_{0}^{+},x_{h}^{+}(x_{0}^{+}))=
\frac {2m} {\lambda} - \frac {\kappa} {2} \log(1+ \frac {4m}
{\kappa\lambda})  \label{eq:ENTRO}
\end{equation}
This expression is positive for all positive values of m. For large
 values of
$\frac m {\kappa\lambda}$ it increases arbitrarily, and as
 $\frac m {\kappa\lambda}$
tends to zero, it goes to zero.
For fixed $\frac m {\kappa\lambda}$, but with the number $\kappa$ of
matter fields
going to infinity, the entropy change diverges.

This result depends upon choosing the right boundary conditions for
 the
Z field, the correct vacuum state. Myers\cite{MY} obtains equivalent
 results by
considering
the appropriate conformal transformation (to the flat asymptotic
vacuum
coordinates) on the $\rho_{h}$ term which appears in
his entropy formula. This adds precisely the contribution to the
 entropy
which
the function $\xi$ gives in the Z-field analysis used here.

To see that the entropy is always increasing for these shockwave
solutions as
the
advanced time increases on the horizons, one can simply write the
entropy
as a function of $x^{+}_{h}$, and finds that,
\begin{equation}
S_{AH}=\frac {\kappa} 2 + \frac {2m} {\lambda} + \frac {\kappa} 2
\log(\frac
{\kappa} 4 +\frac {mx_{h}^{+}} {\lambda x_{0}^{+}})   \label{eq:SAH}
\end{equation}
and
\begin{equation}
S_{EH}=\frac {4m} {\lambda} + \frac {\kappa} 2 \log(\frac
{mx_{h}^{+}} {\lambda x_{0}^{+}(e^{\frac {4m} {\kappa\lambda}}-1)})
+\frac {2mx_{h}^{+}} {\lambda x_{0}^{+}(e^{\frac {4m}
{\kappa\lambda}}-1)}
  \label{eq:SEH}
\end{equation}
both of which clearly increase with advanced time.

This is a second law for these solutions on both the apparent and
event
horizons.
In\cite{MY} a second law is proved to exist more generally
({\it i.e.} all solutions
(\ref{eq:GEN})) for this model.

\section{Conclusion}

Now that there is a semi-classical model which describes black holes
which
evaporate, and which is simple enough to yield exact solutions,
 though not
quite in the simplest and most directly handlable way, the
immediate
task is to extend the classical theory thermodynamics of black
holes,
to cover the new
dynamical situations of the recent models. This is being done,
and has been applied here for the local, well-posed RST model.

Consider an empty spacetime, whose entropy at some
advanced time is  normalised to zero, say. Send in matter at the
speed
of light at this time forming a black hole, and attribute to it
 an intrinsic
entropy. At a later advanced time, the black hole ceases to be, but
the
quantity
we called the entropy has risen above zero, and to what
extent it has done
so depends on how energetic was the original pulse of energy,
the number of degrees of freedom we allow for the carriage of
radiation to infinity, and the model with which we are working.
Once the black hole has disappeared, we are presumably back to the
same
spacetime geometry that we began with, but the entropy  has
increased. We need to know how this quantity is relevant to
information loss.
   An increase in entropy is normally taken to be associated with a
loss of information\cite{FPST}.

Properly defined, the entropy corresponding to `missing information'
for matter in a pure state, which is finely monitored, should
remain constant, although in {\it practice} this monitoring is never
possible, and the coarse-graining then allows energy to degrade.
If there is a black hole, formed from pure-state matter, and no
amount of monitoring could prevent this entropy increasing in the
course of evaporation, then information is said to be lost in
{\it principle}. That is, no one to one map exists between the
initial state preparation and future infinity after the black hole
has disappeared.

In further work, we shall consider the question of how to quantify
the
 loss of
information which appears to occur in the semi-classical black hole
 theory.
What seems needed is a clarifying study of information theory
uesful
 to black hole physicists. In what relation does `information'
stand to the `entropy' of black hole mechanics. This
latter
can be related to the scattering properties of black holes-which are
themselves
 rather difficult concepts-and can be derived directly from the
 action,
the
starting point for most two dimensional models. This work will
be the third
of
three linked papers of which this is the second and \cite{SHM},
which considered
the
general theory of scattering and coherence in two dimensional
semi-classical
theory, was the first.

In Bekenstein's work\cite{BekA}, formulae which
relate the rate of
black hole intrinsic entropy change directly to rate of information
 retrieval are used.
That is,  the deviations from thermality of the radiation from a
 black hole are directly proportional to the information retrieval
rate. It is known that there are such deviations, from the work on
two-dimensional
models\cite{MATHUR,JH4}. Thus if we can determine what these
correspond  to, then we can see whether the black hole can re-emit
all its  information, at least at in the approximation in which our
present notion of spacetime geometry is unchanged. It seems,
however that the information stored in these non-thermal modes is
limited\cite{JH4}.

\section{Acknowledgements}

I thank Stephen Hawking for help and
encouragement.

\end{document}